# Human Luteinizing Hormone and Chorionic Gonadotropin Display Biased Agonism at the LH and LH/CG Receptors

Laura Riccetti[1], Romain Yvinec[2], Danièle Klett[2], Nathalie Gallay[2], Yves Combarnous[2], Eric Reiter[2,*], Manuela Simoni[1,4,5], Livio Casarini[1,4,*], and Mohammed Akli Ayoub[2,3,6,*]

[1] Unit of Endocrinology, Department of Biomedical, Metabolic and Neural Sciences, University of Modena and Reggio Emilia, Modena, Italy.

[2] PRC, INRA, CNRS, Université François Rabelais-Tours, 37380 Nouzilly, France.

[3] LE STUDIUM® Loire Valley Institute for Advanced Studies, 45000 Orléans, France.

[4] Center for Genome Research, University of Modena and Reggio Emilia, Modena, Italy.

[5] Azienda, Ospedaliero-Universitaria di Modena, Modena, Italy.

[6] Biology Department, College of Science, United Arab Emirates University, Al Ain, United Arab Emirates.

*To whom correspondence should be addressed:

Mohammed Akli Ayoub - Biology Department, College of Science, United Arab Emirates University, Al Ain – PO Box 15551, United Arab Emirates - Email: mayoub@uaeu.ac.ae

Livio Casarini - Unit of Endocrinology, Department of Biomedical, Metabolic and Neural Sciences, University of Modena and Reggio Emilia, via Pietro Giardini 1355 - 41126 Modena, Italy - Email: livio.casarini@unimore.it

Eric Reiter, UMR PRC, 37380, Nouzilly, France - Email: Eric.Reiter@inra.fr

**Abstract**

Human luteinizing hormone (LH) and chorionic gonadotropin (hCG) have been considered biologically equivalent because of their structural similarities and their binding to the same receptor; the LH/CGR. However, accumulating evidence suggest that LH/CGR differentially responds to the two hormones triggering differential intracellular signaling and steroidogenesis. The mechanistic basis of such differential responses remains mostly unknown. Here, we compared the abilities of recombinant rhLH and rhCG to elicit cAMP, β-arrestin 2 activation, and steroidogenesis in HEK293 cells and mouse Leydig tumor cells (mLTC-1). For this, BRET and FRET technologies were used allowing quantitative analyses of hormone activities in real-time and in living cells. Our data indicate that rhLH and rhCG differentially promote cell responses mediated by LH/CGR revealing interesting divergences in their potencies, efficacies and kinetics: rhCG was more potent than rhLH in both HEK293 and mLTC-1 cells. Interestingly, partial effects of rhLH were found on β-arrestin recruitment and on progesterone production compared to rhCG. Such a link was further supported by knockdown experiments. These pharmacological differences demonstrate that rhLH and rhCG act as natural biased agonists. The discovery of novel mechanisms associated with gonadotropin-specific action may ultimately help improve and personalize assisted reproduction technologies.

**Introduction**

Luteinizing hormone (LH) and human chorionic gonadotropin (hCG) are two heterodimeric glycoprotein hormones playing key roles in human reproduction. They are produced by the pituitary gland (LH) and placenta (hCG) and circulate as mixtures of differentially glycosylated isoforms which present different half-lives and bioactivities [1-3]. Both hCG and hLH bind to luteinizing hormone/choriogonadotropin hormone receptor (LH/CGR) in human and LHR in non-human species which are mainly expressed in the ovary and testis. In normal female cycle, LH is involved in late follicular maturation and ovulation and it also triggers *corpus luteum* steroidogenic activity. hCG is responsible for maintaining steroidogenic activity by *corpus luteum* over the first four months of pregnancy in women. hCG also inhibits LH and FSH secretion and triggers steroidogenesis in fetal gonads [4,5]. LH/CGR belongs to a subgroup of class A (rhodopsin-like) GPCRs characterized by the presence of multiple leucine-rich repeats (LRRs) in their extracellular amino-terminal domain. Regarding its signaling, LH/CGR is known to mediate the canonical G protein-mediated signaling pathway through coupling to heterotrimeric Gαs protein which activates adenylate cyclase. It results in cAMP accumulation and activation of protein kinase A (PKA), as well as in the exchange protein directly activated by cAMP (EPAC). This, in turn, triggers the activation of multiple downstream kinases that modulate the nuclear activity of cAMP response element-binding protein (CREB) and the expression of the genes involved in the physiological responses to these hormones. Besides, LH/CGR was one of the first GPCRs shown to independently activate two G proteins, leading to both adenylyl cyclase and phospholipase C activation through functional coupling to Gαs and Gαq, respectively [6,7]. More recently, LH/CGR has been also reported to engage a multiplicity of G protein-independent pathways [8] including β-arrestin-dependent pathways [9-11]. For many years, β-arrestins have been considered exclusively as silencers of the GPCRs signaling, prompting ligand-induced receptor internalization and trafficking [12,13]. Now it is well recognized that β-arrestins regulate GPCR signaling and trafficking and are also able to engage G protein-independent signaling, the most studied of which being ERK1/2 pathway [12-14]. The implication of β-arrestins in trafficking and signaling of gonadotropin receptors has also been reported [11,15-21].

For many years, LH and hCG have been assumed to be equivalent, even though distinct physiological [5,22], molecular [23,24] and pharmacological [25,26] features were described [3,27]. Importantly, the phenomenon of biased signaling implies that the binding of a given ligand can stabilize a subset of activated conformations of the receptor thereby leading to a selective modulation of downstream signaling pathways. This has been reported for numerous GPCRs [14,28-30]. In line with this emerging concept, it has been recently proposed that LH and hCG produced as multiple glycosylated isoforms could potentially trigger selective transduction mechanisms at the LH/CGR [1,8]. Recent lines of evidence support this hypothesis, showing that, although their structures are similar and they shared the same receptor, LH and hCG elicit divergent signaling in several cell models [27]. A genomic deletion resulting in the complete absence of exon 10 of LH/CGR found in a hypogonadic patient with type II Leydig cell hypoplasia [31] impaired LH-, but not hCG-induced cAMP production, without affecting ligand binding [32]. More recently, it has been reported in human granulosa cells that hCG displays higher potency than LH on the cAMP/PKA pathway as well as on steroidogenesis, whereas LH is more potent than hCG on ERK1/2 and Akt phosphorylation as well as on related gene expression [33]. The proliferative and steroidogenic/pro-apoptotic effects mediated by LH and hCG *in vitro*, respectively, are amplified by FSH co-treatment [34]. In fact, in goat ovarian granulosa cells, prolonged LH treatment promotes growth and proliferation whereas hCG leads to higher levels of cAMP and decreased proliferation [35].

Despite these recent advances, the potential of hCG and LH to differentially activate Gαs and β-arrestin-dependent pathways at the LH/CGR and LHR has not been evaluated. In addition, the relative contributions of these transduction mechanisms to steroidogenesis are still unknown. In the present study, we used a series of bioluminescence and time-resolved resonance energy transfer (BRET and TR-FRET) and reporter assays to quantitatively assess the signaling promoted by rhCG and rhLH in real-time and living cells as previously reported [10,11]. The canonical Gαs/cAMP pathway as well as β-arrestin 2 recruitment were analyzed upon the activation of human LH/CGR transiently expressed in HEK293 cells or LHR endogenously expressed in murine Leydig tumor cell line (mLTC-1) [11,36]. In addition, the degree of bias between rhCG- and rhLH-mediated responses was quantified using the operational model of Black and Leff [37] and previously detailed procedures [11,38].

**Materials and Methods**

**Recombinant gonadotropins:** Human recombinant LH (Luveris®) and rhCG (Ovitrelle®) were kindly provided by Merck KGaA (Darmstadt, Germany). Molar concentrations of the recombinant rhCG and rhLH were calculated based on their respective specific activities (6500 and 5360 IU) and molecular weights (37 and 30 kDa).

**Cell Culture and Transfection:** HEK293 cells were grown in complete DMEM medium supplemented with 10% (v/v) fetal bovine serum, 4.5 g/l glucose, 100 U/ml penicillin, 0.1 mg/ml streptomycin, and 1 mM glutamine (Invitrogen, Carlsbad, CA, USA). mLTC-1 cells (ATCC CRL-2065, LCG Standards, Molsheim, France) were grown in complete RPMI medium supplemented with 10% (v/v) fetal bovine serum, 50 µg/ml gentamicin, 10 units/ml penicillin and 10µg/ml streptomycin. Transient transfections were performed by reverse transfection in 96-well plate using Metafectene PRO (Biontex, München, Germany) following the manufacturer's protocol and using $10^5$ cells, 100 ng of total plasmids and 0.5 µl of Metafectene PRO per well.

**Small interfering RNA transfection:** The siRNA sequence 5'-AAAGCCUUCUGUGCUGAGAAC-3' was used to target mouse β-arrestin 1 (position 439-459 relative to the start codon) whereas sequence 5'-AAACCUGUGCCUUCCGCUAUG-3' was used to target mouse β-arrestin 2 (position 175-193 relative to the start codon) [39]. One small RNA duplex with no silencing effect was used as a control (5'-UUCUCCGAACGUGUCACGU-3'). The siRNAs were synthesized by GE Healthcare Dharmacon (Velizy-Villacoublay, France). Early passage mLTC-1 cells at 30% confluency in 100 mm dishes were transiently transfected with GeneSilencer following the manufacturer's recommendations (Genlantis, San Diego, CA, USA). Forty-eight hours after transfection, cells were seeded into assay plates. All assays were performed three days after transfection.

**BRET sensors**: In order to measure cAMP real-time response in living cells, HEK293 cells were transiently transfected with two plasmids coding for hLH/CGR (kindly provided by A. Ulloa-Aguirre, Universidad Nacional Autónoma de México, México) and the BRET-based cAMP sensor CAMYEL (kindly provided by L.I. Jiang, University of Texas, Texas, USA) as previously reported [10,11]. For mLTC-1 cells, only CAMYEL plasmid was transfected since they naturally express the endogenous LHR.

For the assessment of β-arrestin 2 recruitment, HEK293 cells were transiently co-transfected with plasmids coding for hLH/CGR C-terminally fused to the BRET donor Rluc8 (kindly provided by Dr. A. Hanyaloglu, Imperial College, London, UK) and for β-arrestin 2 N-terminally fused to the BRET acceptor yPET (kindly provided by Dr. M.G. Scott, Cochin Institute, Paris, France). Upon rhCG and rhLH stimulation, β-arrestin 2 translocates to the receptor, leading to energy transfer between Rluc8 and yPET, and as a consequence, to dose-dependent increases in BRET signals. The conformational rearrangements within β-arrestin 2 were measured using β-arrestin 2 double brilliance sensor [40-42]. For this, HEK293 cells were transiently co-transfected with the plasmids coding for hLH/CGR and for Rluc8-β-arrestin 2-RGFP fusion protein (kindly provided by Dr. R. Jockers, Cochin Institute, Paris, France). Then, changes in intramolecular BRET signals within sensor were monitored upon cell stimulation with increasing doses of rhCG and rhLH.

**Cell stimulations and BRET measurements:** For the end-point dose-response experiments, medium was aspirated and cells were re-suspended in 40 µl/well of PBS 1X, 1 mM HEPES. Cells were incubated for 30 minutes at 37°C in a total volume of 40 µl/well of PBS 1X, 1 mM HEPES containing or not increasing concentrations of rhCG and rhLH. BRET measurements were performed upon addition of 10 µl/well of 5 µM Coelenterazine h (Interchim, Montluçon, France), using Mithras LB 943 plate reader (Berthold Technologies GmbH & Co. Wildbad, Germany). For the real-time kinetics, cells were re-suspended in 60 µl/well of PBS 1X, 1 mM HEPES and, for cAMP kinetics, 200 µM IBMX. BRET measurement was immediately performed upon addition of 20 µl/well of $EC_{50}$ concentrations of rhCG and rhLH as previously calculated in dose-response experiments, and 20 µl/well of coelenterazine h.

*Cre*-dependent reporter assay: HEK293 cells were transiently transfected with hLH/CGR and the pSOM-Luc plasmid expressing the firefly luciferase reporter gene under the control of the cAMP Responsive Element of the somatostatin promoter region [43]. mLTC-1 cells were transiently transfected with the pSOM-Luc plasmid alone. After 48-hours, cells were split into 96-well plates. The day after, cells were stimulated 6 hours with increasing doses of hormones, then washed twice with ice-cold PBS and lysed in 200 µl of lysis buffer (Promega, Madison, WI, USA). Luciferase activity was measured using the luciferase assay system supplied by Promega. An aliquot (20 µl) of each sample was mixed with 50 µl of luciferase assay reagent and the emitted light was measured in Mithras LB 943 plate reader. Values were expressed in relative luciferase activity units (RLU).

**Steroid measurements:** Progesterone and testosterone levels were measured in the supernatants of mLTC-1 cells cultured in complete RPMI medium. For progesterone, cells were first seeded in 48-well plates ($10^5$ cells/well) for three days and then re-suspended in serum-free RPMI for one hour and stimulated or not with increasing doses of rhCG and rhLH for 3 hours. Cell supernatants were collected and stored at -20°C until analysis. Progesterone production was measured with a home-made competitive ELISA assay [44]. Briefly, a 96-wells plate was coated overnight at 4°C with a goat anti-mouse IgG antibody, 10 ng/well (UP462140, Interchim, Montluçon, France). After three washes in PBS 1X containing 0.1% Tween 20, non-specific sites were saturated 1 hour with 200 µl/well of PBS-Tween 20 supplemented with 0.2% BSA. Standard progesterone (Q2600, Steraloids, USA) in PBS-Tween 20-BSA or mLTC-1 cells supernatants (25µl per well of 1:50 or 1:100 dilution) were then plated on the empty plate. Progesterone-11-Hemisuccinate-HRP (Interchim) were added, together with 36 ng/well of mouse anti-P4 antibody (AbD Serotec, Biogenesis, Interchim). The plate was incubated for 4 hours at room temperature, washed and 100 µl/well of TMB ELISA substrate standard solution (Interchim) was added. The mixture was incubated for 20 min at room temperature in the dark. The reaction was stopped with 2N $H_2SO_4$ and absorbance was measured at 450 nm using Sunrise™ absorbance reader (Tecan, France).

Testosterone levels were measured by using an HTRF®-based assay kit (CisBio Bioassays, Codolet, France) following the manufacturer's protocol and as previously described [11]. Briefly, cells were seeded in 96-well plate ($10^4$ cells/well), starved overnight in serum-free RPMI medium and incubated 3 hours at 37°C in 100 µL/well of serum-free RMPI medium containing or not increasing concentrations of rhCG and rhLH. Then 10 µl of culture supernatants were transferred into 384-well and mixed with 10 µl of a mixture of HTRF mix containing an anti-testosterone antibody and testosterone labeled with Terbium and d2 fluorophores, respectively. TR-FRET signals were detected using Mithras LB 943 plate reader (Berthold Technologies GmbH & Co. Wildbad, Germany).

**cAMP HTRF-based assay:** Intracellular cAMP levels were measured using a homogeneous time-resolved fluorescence (HTRF®) assay kit (CisBio Bioassays) [10,11] according to the manufacturer's instructions.

**Data analysis**: BRET data were represented as "Induced BRET Changes" by subtracting the 540 nm/480 nm ratio of the non-treated cells from the same ratio of cells stimulated with increasing doses of rhCG and rhLH. The % of responses in BRET and steroid measurement were obtained by taking as 100% the maximal response of rhCG at 10 nM in the different assays. All the results were fitted following the appropriate nonlinear regression equations using GraphPad Prism software (San Diego, CA, USA). Statistical analyses were performed with unpaired t-test (for the $E_{max}$ and $EC_{50}$ values) in HEK293 and mLTC-1 cells and Two-way ANOVA (for siRNA β-arrestin data in mLTC-1 cells), both using the GraphPad Prism software.

**Bias calculation**: The bias factor (B.F.) was obtained after statistical fitting of the data to the operational model reported by Black and Leff [37], as previously reported [11].

**Results**

**LH/CGR-promoted cAMP Response:** To compare the respective efficacies and potencies of rhCG and rhLH, we first assessed the ability of either hormone to elicit the accumulation of cAMP, the prototypical second messenger produced upon coupling and activation of the heterotrimeric Gαs protein by LH/CGR. For this purpose, HEK293 cells transiently co-expressing human LH/CGR and the BRET-based cAMP sensor (CAMYEL) were used as previously reported [10,11]. Changes in BRET signal were monitored in living cells after 30 minutes of incubation with increasing doses of rhCG and rhLH. As expected, both hormones showed very potent effects on the cAMP signaling pathway with $EC_{50}$ values in the pM range as previously shown [10,11] (**Fig.1A**)(**Table 1**). Consistent with previous reports, the rhCG dose-response curve is significantly shifted towards the lower doses compared to that of rhLH [33]. Indeed, the $EC_{50}$ values of rhCG response was found approximately 16 times lower than that of rhLH ($p<0.0001; n=7$)(**Table 1**). By contrast, at the saturating doses both hormones promoted similar maximal cAMP responses indicating that both act as full agonists on LH/CGR-promoted cAMP production. We also investigated cAMP accumulation in real-time kinetics by stimulating cells with different concentrations (0.05, 0.5, and 5 nM) of rhCG and rhLH to cover the $EC_{50}$ and sub-maximal doses according to the dose curves shown in **Fig.1A**. As shown in **Fig.1B**, a relationship between the doses and the kinetics was observed with faster and saturating responses obtained when increasing doses were used. Indeed, when rhCG and rhLH were compared at their respective sub-saturating doses (*i.e.*: 0.01 nM for rhCG and 0.5 nM for rhLH), both promoted a rapid cAMP response with no difference in their kinetics ($t_{1/2}$ around 1 min) (**Fig.1B**). The response reached a plateau even after 30 minutes of stimulation (**Fig.1B**) and this cannot be due to a saturation of the cAMP response in our system. In fact, at a higher and saturating dose of both hormones (5 nM), the cAMP response was further increased with faster ($t_{1/2}$ value around 0.5 min) and equivalent kinetics for both hormones (**Fig.1B**).

To investigate the cAMP pathway on endogenously expressed LHR, we used a mouse Leydig tumor cell line (mLTC-1)[45] either transfected with CAMYEL sensor (**Fig.1C**) or not (**Fig.1D**), as previously reported [10,11]. Again, BRET measurements revealed rhCG dose-response curve shifted towards the lower doses compared to rhLH (**Fig.1C**). Specifically, the $EC_{50}$ value of rhCG was approximately 6 times lower than that of rhLH ($p=0.0067; n=5$) (**Table 2**). To further confirm these data and to exclude any artifact due to the transfection of CAMYEL sensor in mLTC-1 cells, we also measured cAMP levels in native cells using HTRF-based assay, which does not require any cell transfection. We found a similar pattern of rhCG- and rhLH-induced cAMP responses (**Fig.1D**). It is worth noting that consistent with HEK293 cell data, rhCG and rhLH promoted similar maximal cAMP responses in mLTC-1 cells despite the differences in their potencies confirming that both are full agonists for cAMP pathway (**Fig.1C** and **D**)(**Table 2**).

**β-arrestin 2 recruitment and activation:** β-arrestins are known to play important roles not only in desensitization/internalization of GPCRs but also in their signaling [12-14]. Here, we examined the recruitment of β-arrestin 2 upon exposure to increasing doses of rhCG and rhLH using BRET technology as previously reported [10,11]. HEK293 cells were transiently co-transfected with plasmids coding for hLH/CGR-Rluc8 and for yPET-β-arrestin 2 fusion proteins and BRET measurements were performed in a dose-dependent manner and in living cells. As shown in **Fig.2A**, rhCG and rhLH stimulation significantly promoted β-arrestin 2 recruitment to hLH/CGR in a dose-dependent manner with the $EC_{50}$ value of rhCG 12 times lower than that of rhLH (*p=0.044; n=5*)(**Table 1**). The significant difference in the two hormone potencies is consistent with the data on cAMP pathway (**Fig.1**)(**Table 1**). Low potencies were found for both hormones on β-arrestin recruitment compared to cAMP responses consistent with our previous report indicating that a higher receptor occupancy rate is needed to engage β-arrestin recruitment [10,11]. Interestingly, our data reveal that rhLH seems to exhibit partial agonistic activity compared to rhCG (*p=0.0104; n=5*)(**Fig.2A**)(**Table 1**). Real-time kinetic analysis of BRET signals in cells stimulated with either $EC_{50}$ (10 nM and 50 nM) or saturating (500 nM) concentrations of both rhCG and rhLH, showed a significant difference in β-arrestin 2 recruitment kinetics between the two hormones with $t_{1/2}$ values of *6.15 ± 0.17 min* and *13.32 ± 0.32 min* with 500 nM of rhCG and rhLH, respectively (**Fig.2B**). These kinetic data are in keeping with the partial activity of rhLH compared to rhCG on β-arrestin recruitment even at the saturating doses of the hormone (**Fig.2B**).

Next, we investigated the impact of hLH/CGR activation by rhCG or rhLH on β-arrestin 2 conformations using a previously reported double brilliance β-arrestin 2 BRET sensor [40-42]. In this sensor, β-arrestin 2 is fused with both a BRET donor and acceptor. Therefore, any change in the intramolecular BRET signals reflects a conformational change in β-arrestin 2. Interestingly, real-time kinetic analyses with 0.25 μM of hormones showed stronger induction of conformational changes under LH than rhCG treatment within β-arrestin sensor (**Fig.2C**). However, no significant difference in the half-time (around 12 minutes) between the two hormones was observed (**Fig.2C**). Such a difference between rhCG and rhLH on β-arrestin conformation was observed at different concentrations of the hormones (**Fig.2D**). Together, the data with β-arrestins are consistent with the different sensitivity of LH/CGR to rhCG and rhLH suggesting the existence of hormone-specific receptor conformation which could in turn impact β-arrestin conformation.

**LHR-promoted integrated responses:** To further assess the impact of the differences elicited by the two hormones at the transductional level, we sought to measure downstream read-outs in the signaling pathways. In HEK293 cells, we used a *cre*-dependent reporter gene, pSOM-Luc, as an indicator of LH/CGR-induced transcriptional activation [11,43]. As shown in **Fig.3A**, both hormones showed clear dose-dependent activation of luciferase activity with $EC_{50}$ of rhCG almost 5 times lower than that of rhLH (*p=0.0004; n=3*)(**Table 1**). By contrast, the two hormones showed similar maximal responses expressed in relative luciferase activity (**Fig.3A**)(**Table 1**). These observations are consistent with the cAMP data in HEK293 and mLTC-1 cells as shown in **Fig.1**.

Next, we explored more distal responses to rhCG and rhLH in mLTC-1 cells endogenously expressing LHR. For this, we measured pSOM-Luc activation in *Cre* reporter assay (**Fig.3B**), progesterone (**Fig.3C**) and testosterone (**Fig.3D**) responses upon stimulation with increasing doses of rhCG and rhLH. In the *Cre* reporter assay, we observed a significant shift (*p=0.0202; n=4*) similar to the one observed in HEK293 cells with rhCG being more potent than rhLH, (**Fig.3B**)(**Table 2**). In progesterone assay, we found that the $EC_{50}$ of rhCG was approximately 15 times lower than that of rhLH (*p=0.0011; n=3*)(**Fig.3C**)(**Table 2**). This observation is in agreement with cAMP data (**Fig.1**), β-arrestin data (**Fig.2A**), and *Cre*-luciferase-based cAMP assay (**Fig.3A**) data confirming that rhCG is more potent than rhLH on LH/CGR. Interestingly, the maximal progesterone produced by rhLH was only 49.8 ± 2.9 % (*p<0.0001; n=3*) of that promoted by rhCG, indicating a partial response of rhLH on progesterone synthesis (**Fig.3C**)(**Table 2**). Such a partial response of rhLH is consistent with that observed for LH/CGR-promoted β-arrestin 2 recruitment (**Fig.2A**). In the case of testosterone, we also observed a higher potency of rhCG (≈13 folds) compared to rhLH, (*p=0.0084; n=4*)(**Fig.3D**)(**Table 2**), which is consistent with progesterone data (**Fig.3C**), cAMP responses in HEK293 and mLTC-1 cells (**Fig.1A and C**) as well as β-arrestin in HEK293 cells (**Fig.2A**). We also noticed that both hormones were clearly more potent at activating testosterone than progesterone production (**Table 2**). However, by contrast to progesterone response, there was no significant difference in the maximal testosterone response between the two hormones (*p=0.3848; n=4*)(**Fig.3D**). Together, this illustrates the complexity of the steroidogenic pathways engaged by LHR, suggesting that distinct signaling pathways might control the production of progesterone and testosterone as previously proposed [11].

**Implication of β-arrestins in progesterone production:** The similarity in the partial effects of rhLH on β-arrestin (**Fig.2A**) and progesterone (**Fig.3C**) responses compared to rhCG suggests that β-arrestins may be implicated in progesterone production. Therefore, we examined the contribution of β-arrestin-dependent transduction on the control of LHR-mediated steroidogenesis using siRNA-mediated depletion of endogenous β-arrestin 1 or β-arrestin 2 in mLTC-1 cells. For this, control, β-arrestin 1 or β-arrestin 2-depleted mLTC-1 cells were exposed to increasing doses of rhCG (**Fig.4A, C and E**) and rhLH (**Fig.4B, D and F**). Progesterone and testosterone were measured in parallel in the same cells. Our data confirm the partial progesterone response mediated by rhLH compared to rhCG as shown in **Fig.3B**. However, on testosterone production there was no difference in hormone efficacy but rhCG was more potent that rhLH confirming the data shown in **Fig.3D**. Interestingly, we observed that the depletion of both β-arrestin 1 and β-arrestin 2 led to a partial decrease in the progesterone production induced by rhCG (**Fig.4A**) and rhLH (**Fig.4B**) compared to control siRNA-transfected mLTC-1 cells. The statistical analysis demonstrates the significance of β-arrestin depletion for both rhCG ($p<0.0001$ for siRNA β-arrestin 1 and $p=0.0233$ for siRNA β-arrestin 2; $n=3$) and rhLH ($p=0.001$ for siRNA β-arrestin 1 and $p=0.0275$ for siRNA β-arrestin 2; $n=3$) with siRNA β-arrestin 1 being more efficient. Moreover, β-arrestin depletion showed stronger inhibitory effects on testosterone production mediated by either rhCG (**Fig.4C**) or rhLH (**Fig.4D**) ($p<0.0001$ for siRNA β-arrestin 1 and β-arrestin 2; $n=3$). The specificity of the effects of β-arrestin depletion on steroidogenesis was demonstrated by measuring cAMP production in parallel in the siRNA-transfected mLTC-1 cell samples. As anticipated, the depletion of β-arrestin 1 or 2 had no effect on rhCG- (**Fig.4E**) or rhLH- (**Fig.4F**) mediated cAMP production whatever the dose of the hormones used, thus ruling out toxic or side effects that the siRNA may have exerted on these cells. Together, these data suggest the partial implication of β-arrestins in LHR-mediated progesterone and argue for a significance of β-arrestin bias between rhCG and rhLH in HEK293 cells.

**Calculation of biases between rhCG and rhLH:** Biases were calculated using data-fitting of the operational model [37] and the procedure previously detailed [11,38]. Bias factor combines both efficacy and potency to quantify the imbalance between two cell responses for an agonist, in comparison to a reference agonist on the same receptor and within the same cell model. For this, we used rhCG as the reference ligand for the different assays in both HEK293 and mLTC-1 cells (**Table 3**). A bias exists if the bias factor is significantly different from 1 (unpaired t-test). In HEK293 cells, we found that rhLH, compared to rhCG, was significantly biased towards *cre*-dependent reporter gene against both β-arrestin 2 and cAMP (**Table 3**). To some extent, rhLH is also biased towards β-arrestin 2 against cAMP (**Table 3**). In other words, in HEK293 cells, rhLH preferentially induces: *Cre*-dependent transcription > β-arrestin 2 recruitment > cAMP, compared to rhCG. This finding may be considered counterintuitive since Cre reporter assay and cAMP are connected by the activation of cAMP response element-binding protein (CREB) which is primarily under the control of cAMP/PKA pathway. However, it is also well documented that CREB activation integrates other signaling pathways such as ERK and p38 MAPKs, p90RSK or CAMKs. In that sense, Cre reporter assay can be viewed as a read out which integrates several signalling pathways, not just cAMP/PKA. In mLTC-1 cells, rhLH-promoted response was significantly biased towards cAMP compared to progesterone and testosterone and a moderate bias appeared towards testosterone compared to progesterone (**Table 3**). This means that, in mLTC-1 cells, rhLH preferentially induces: cAMP > testosterone > progesterone, compared to rhCG. Overall, these bias calculations confirm the differences between the pharmacological profiles of rhCG and rhLH on the different responses in HEK293 and mLTC-1 cells.

**Discussion**

In the present work, we investigated the differential activity exerted by rhCG and rhLH upon binding to their common receptor, human LH/CGR transiently expressed in HEK293 cells or mouse LHR endogenously expressed in mLTC-1 cells. The attention was focused on comparing their pharmacological profiles on the heterotrimeric G proteins G$\alpha$s/cAMP pathway as well as β-arrestins and their contribution to signal transduction mechanisms leading to the modulation of steroidogenesis. For cAMP, rhCG was more potent than rhLH but both hormones promoted full activation of the receptor at saturating doses in HEK293 as well as mLTC-1 cells. A similar difference in potency was also observed on β-arrestin recruitment in HEK293 cells, however an interesting difference was also found in hormone efficacy since rhLH promoted only a partial response. We also examined the action of rhLH and rhCG on progesterone and testosterone production in mLTC-1 cells confirming the higher potency of rhCG compared to rhLH. Interestingly, we found that rhLH exhibited a partial activity for progesterone compared to rhCG while both fully promoted testosterone production. Altogether, our data show that rhCG is more potent than rhLH on LH/CGR activity in HEK293 and mLTC-1 cells which is consistent with previous studies in COS-7 and granulosa cells [33]. Moreover, the partial agonism of rhLH on LH/CGR on both β-arrestin recruitment and progesterone production suggests a link between these two events. It can be observed that the minimal doses of rhCG or rhLH promoting full testosterone responses trigger only very partial (20-25%) progesterone response. It reflects the physiological role of hCG, which serves to induce progesterone synthesis, while LH is a mediator of testosterone production. Indeed, in Leydig cells, androgen synthesis occurs mainly *via* the so-called Δ5-pathway, while progesterone is a "parallel accumulation product" falling within the relatively ineffective Δ4-pathway [46]. We could speculate that, differently to hCG, LH is fully active on the Δ5- rather than Δ4-pathway. The results provided by β-arrestins silencing by siRNA (**Fig.4**) agree with this view, since these molecules are required to fully support LH/hCG-mediated testosterone but not progesterone synthesis.

In the present study, we are using immortalized cell lines as model systems. It is important to determine to what extent the data obtained in these cells can be indicative of the physiological mechanisms that occur in the ovary. In fact, the transduction mechanisms occurring directly downstream of the ligand-LHCGR complex, such as G$_{\alpha s}$ coupling and cAMP production or β-arrestins recruitment, have been reported to be mostly reproducible across a wide range of cell models. This is corroborated by previous observation in human and goat primary granulosa cells [33-35], in transfected COS-7 and human immortalized granulosa cell lines [33], as well as in mouse primary Leydig cells [47]. In all these cell systems, hCG treatment results in higher cAMP production than LH. One can rationalize that these mechanisms largely rely on the ligand and receptor amino acid sequence, conformational changes, etc., all properties that are nearly identical in granulosa and Leydig cells, as well as in transfected cell lines. Indeed, we demonstrated that hCG is more potent than LH, in terms of cAMP production, in all these cell models. The amino-acid changes existing between the rodent and the human receptor have been shown to have minimal, if any, effects hLH or hCG binding rates [23]. According to this principle, downstream outputs, which are located farer from the ligand-receptor complex, may lead to more cell type-dependent variations. This is indeed the case when comparing the mLTC-1 cell line used in the present paper and the primary mouse Leydig cells, which were used in another recently published study [47]. In primary Leydig cells, we demonstrated that LH and hCG treatment results in different cAMP production but equal testosterone dose-response curves, whereas in mLTC-1 cells, hCG is more potent than hLH at inducing testosterone production. Taken together, these findings indicate that hCG has a higher steroidogenic potential than LH, even if, in physiological settings, their effects may differ as a consequence of tissue-specific modulatory events.

From the molecular point of view, LH- and hCG-induced signaling is known to be differently modulated by exon 10 deletion in LH/CGR, which results in structural and spatial rearrangements at the hinge region of the receptor [32]. In the presence of this deletion, LH signaling is impaired while hCG signaling remain unchanged, suggesting divergences between hCG- and LH-receptor interactions and actions. In addition, hCG and LH were recently shown to interact differently with the hinge region of the receptor [25] and only hCG is capable of inducing both cis- and trans-activation of human LH/CGR [26]. Moreover, both hCG and LH were reported to have similar association rate on rat LHR ($3 \times 10^8$ $M^{-1}.min^{-1}$) but different dissociation rate (25 h for hCG and ~9 h for hLH) constants indicating higher residence time of hCG on its receptor [23]. Such a finding was recently supported by dissociation assays on gonadotropin-promoted cAMP production in mLTC-1, suggesting weaker dissociation of hCG from LHR compared to hLH [36]. Furthermore, our real-time cAMP measurements clearly showed differences in the kinetics of the two hormones with rhCG inducing faster cAMP responses than rhLH. This indicates the relationship between the difference in hormone potencies and the kinetics of their induced responses, suggesting a link between pharmacological bias and response kinetics as recently reported [48]. Collectively, these observations suggest that the slower dissociation rate of rhCG compared to that of rhLH and/or the differences in the conformations of hormone-receptor complexes may explain the differences between rhCG- and rhLH-promoted responses.

Our quantitative pharmacological profiling revealed striking peculiarities when comparing the maximal responses elicited by the two gonadotropins on the different readouts. Indeed, even though identical maximal cAMP responses were reached with either gonadotropin, whatever the cell type (HEK293 and mLTC-1), rhLH led to significantly weaker maximal β-arrestin 2 recruitment and progesterone production than rhCG. Thus, it could be hypothesized that β-arrestin is a limiting factor for maximal progesterone production. In other words, rhLH like rhCG is full agonist on LH/CGR for cAMP and testosterone production, whereas it is only partial agonist for β-arrestin recruitment and progesterone production. The fact that rhLH was full agonist on testosterone and had only a partial efficacy on progesterone further demonstrates that the full action of the hormone, in Leydig cells, is exerted *via* Δ5-pathway, through 17-OH-pregnenolone production (instead of progesterone) as a precursor of testosterone. This is also supported by our recent study using two FSHR and LHR negative allosteric modulators (NAMs) showing differential antagonism of the two NAMs on progesterone and testosterone [11]. Moreover, we observed that both hormones differently affected the conformation of β-arrestin 2 as assessed in double brilliance BRET assay used as conformational sensor of β-arrestins [40,42]. In this assay, rhCG elicited less BRET changes within the β-arrestin 2 sensor than rhLH which is consistent with the idea that the two hormones stabilize different conformations of β-arrestin as shown for other GPCRs depending on the ligands applied [41,49,50].

Together, these observations strongly suggest that rhCG and LH differentially recruit ß-arrestins and activate downstream pathways which may control progesterone production. We hypothesized that β-arrestin-dependent transduction could be involved in the control of the balance between progesterone and testosterone production as recently reported [11]. Supporting this view, we demonstrated that depletion of endogenous β-arrestins in mLTC-1 cells using selective siRNA led indeed to partially but significantly reduced progesterone production. This supports the link between the partial agonism of rhLH on both β-arrestin and progesterone production. In contrast, such depletion had no effect on hormone-promoted cAMP production in mLTC-1 excluding a role of β-arrestins in cAMP production as expected. Altogether, our results support to the concept of biased agonism exerted by rhCG and rhLH and bear the notion that LH/CGR can discriminate the binding of the two hormones, thereby triggering different transduction mechanisms hence intracellular responses. In addition to the pharmacological profiling of rhCG and rhLH, our study and the bias analysis reveal the importance to quantify the balance between the different signaling pathways for each hormone. The fact that both hormones naturally coexist but to different extent and at different stages during folliculogenesis and pregnancy, raises intriguing prospects. The use of these hormones in assisted reproduction could also be impacted by the present findings. Finally, the impact of our study goes beyond the gonadotropin receptors and reproduction since it describes an interesting example of biased signaling involving two endogenous hormones activating a common GPCR. This further demonstrates the physiological relevance of the signaling bias at GPCRs with strong impact on our understanding of GPCR pharmacology and signaling and potential applications in drug discovery programs [51,52].

**Acknowledgements**

This work has been funded by Région Centre ARD2020 Biomédicaments. LR is funded by Associazione Scientifica in Endocrinologia, Andrologia e Metabolismo and Erasmus+ Grant. MAA is funded by LE STUDIUM® Loire Valley Institute for Advanced Studies and AgreenSkills Plus.

**Disclosure Statement**

The authors have nothing to disclose.

**Author contribution statement**

"L.R. and M.A.A. designed and performed most of the BRET and FRET experiments, wrote the main manuscript text, and prepared the figures and tables; R.Y. performed bias calculations; D.K and Y.C. contributed to steroid measurements; N.G. contributed to siRNA experiments; E.R., M.S., and L.C. designed the experiments and wrote the manuscript. All authors reviewed the manuscript."

**Table 1: Efficacy ($E_{max}$) and efficiency ($EC_{50}$) of rhCG and rhLH on LH/CGR measured in different assays in HEK293 cells.** The $E_{max}$ values are represented as hormone-induced BRET changes (for cAMP and ß-arrestin 2) and relative luciferase activity x $10^3$ (for *Cre*-reporter assay). Statistical analyses were performed with unpaired t-test (*** $p < 0.001$; ** $p < 0.01$; * $p < 0.05$, ns non-significant).

| Responses | rhCG | | | rhLH | | |
|---|---|---|---|---|---|---|
| | $E_{max}$ | $EC_{50}$ | n | $E_{max}$ | $EC_{50}$ | n |
| cAMP | 0.43 ± 0.02 | 17.64 ± 6.97 pM | 8 | 0.42 ± 0.01[ns] | 286.49 ± 50.29 pM[***] | 8 |
| ß-arrestin 2 | 0.21 ± 0.01 | 10.26 ± 2.50 nM | 5 | 0.14 ± 0.01[**] | 129.10 ± 69.96 nM[*] | 5 |
| *Cre*-reporter assay | 222.12 ± 9.64 | 37.47 ± 4.94 pM | 3 | 214.03 ± 9.58[ns] | 215.45 ± 20.81 pM[***] | 3 |

**Table 2: Efficacy ($E_{max}$) and efficiency ($EC_{50}$) of rhCG and rhLH on LHR measured in different assays in mLTC-1 cells.** The $E_{max}$ values are represented as hormone-induced BRET changes (for cAMP by BRET), % of rhCG-induced response (for cAMP by HTRF, progesterone and testosterone) and relative luciferase activity (for *Cre*-reporter assay). Statistical analyses were performed with unpaired t-test (*** $p < 0.001$; ** $p < 0.01$; * $p < 0.05$, ns non-significant).

| Responses | rhCG | | | rhLH | | |
|---|---|---|---|---|---|---|
| | $E_{max}$ | $EC_{50}$ | n | $E_{max}$ | $EC_{50}$ | n |
| cAMP (BRET) | 0.45 ± 0.02 | 68.82 ± 22.30 pM | 5 | 0.44 ± 0.02[ns] | 459 ± 105.35 pM[**] | 5 |
| cAMP (HTRF) | 100% | 97.37 ± 47.15 pM | 3 | 102 ± 5.8%[ns] | 1980.33 ± 868.64 pM[*] | 3 |
| *Cre*-reporter assay | 1910 ± 111 | 24.97 ± 16.20 pM | 4 | 1971 ± 168[ns] | 317.62 ± 210.70 pM[*] | 4 |
| Progesterone | 100% | 19.72 ± 4.28 pM | 3 | 49.78 ± 2.90%[***] | 305.83 ± 25.22 pM[**] | 3 |
| Testosterone | 100% | 2.19 ± 0.91 pM | 4 | 93.33 ± 6.5%[ns] | 29.05 ± 6.90 pM[**] | 4 |

**Table 3: Bias factor on the different responses in HEK293 and mLTC-1 cells.** The bias factor values of each response pair reflect the biased agonism of rhLH towards the first response compared to the second, all in comparison to rhCG taken as the reference hormone.

| HEK293 cells | Bias factor | mLTC-1 cells | Bias factor |
|---|---|---|---|
| ß-arrestin 2 / cAMP | **1.81** (*p=0.0655*) | cAMP / Progesterone | **5.25** (*p=0.0034*) |
| *Cre*-reporter assay / cAMP | **4.84** (*p=0.0000177*) | cAMP / Testosterone | **2.39** (*p=0.0301*) |
| *Cre*-reporter assay / ß-arrestin 2 | **2.67** (*p=0.00988*) | Testosterone / Progesterone | **2.19** (*p=0.0617*) |

**Figure Legends**

**Figure 1: LH/CGR-mediated cAMP production**. cAMP responses induced by rhCG and rhLH in HEK293 cells transiently co-transfected with hLH/CGR and CAMYEL sensor (**A**, **B**) and in mLTC-1 cells transiently transfected with CAMYEL sensor alone (**C**) or not (**D**). Cells were stimulated 30 minutes with increasing doses (**A**, **C** and **D**) or immediately with the indicated doses (**B**) of rhCG and rhLH, before cAMP production was measured by BRET (**A**, **B** and **C**) or HTRF (**D**). Data are means ± SEM of 3-8 independent experiments. The kinetic curves in panel **B** are representative of 3 experiments performed in triplicate.

**Figure 2: LH/CGR-promoted β-arrestin 2 recruitment and activation in HEK293 cells**. HEK293 cells transiently co-expressing either hLH/CGR-Rluc8 and yPET-β-arrestin 2 (**A** and **B**) or wildtype hLHR/CGR and Rluc8-β-arrestin 2-RGFP (**C**) were stimulated 30 minutes with increasing doses (**A** and **D**) or immediately with the indicated doses (**B**) or 0.25 μM (**C**) of rhCG and rhLH before BRET was measured. Data are means ± SEM of 3-5 independent experiments. The kinetic curves in panel **B** are representative of 3 experiments performed in triplicate.

**Figure 3: LH/CGR-promoted integrated responses in HEK293 and mLTC-1 cells**. **A**) HEK293 transiently co-transfected with plasmids coding for LH/CGR and pSOM-Luc reporter gene were stimulated 6 hours with increasing doses of rhCG and rhLH before luciferase activity was measured (**A**). mLTC-1 cells transiently transfected with plasmid coding pSOM-Luc reporter gene were stimulated 6 hours with increasing doses of rhCG and rhLH before luciferase activity was measured (**B**). For LHR-mediated steroid production, mLTC-1 cells were stimulated 3 hours with increasing doses of rhCG and rhLH and the supernatant levels of progesterone (**C**) and testosterone (**D**) were measured by ELISA or HTRF, respectively. Data are means ± SEM of 3-5 independent experiments performed in single or duplicate.

**Figure 4: Implication of β-arrestins in LH/CGR-promoted steroid production in mLTC-1 cells**. mLTC-1 cells were transiently transfected with either control, β-arrestin 1 or β-arrestin 2 siRNAs. Then, cells were stimulated 3 hours with increasing doses of rhCG (**A**, **C** and **E**) and rhLH (**B**, **D** and **F**) before progesterone (**A** and **B**), testosterone (**C** and **D**) and cAMP (**E** and **F**) productions were measured. The maximal rhCG-mediated responses obtained at 10 nM with control siRNA was taken as 100%. Data are means ± SEM of 3 independent experiments performed in duplicate.

# Figure 1

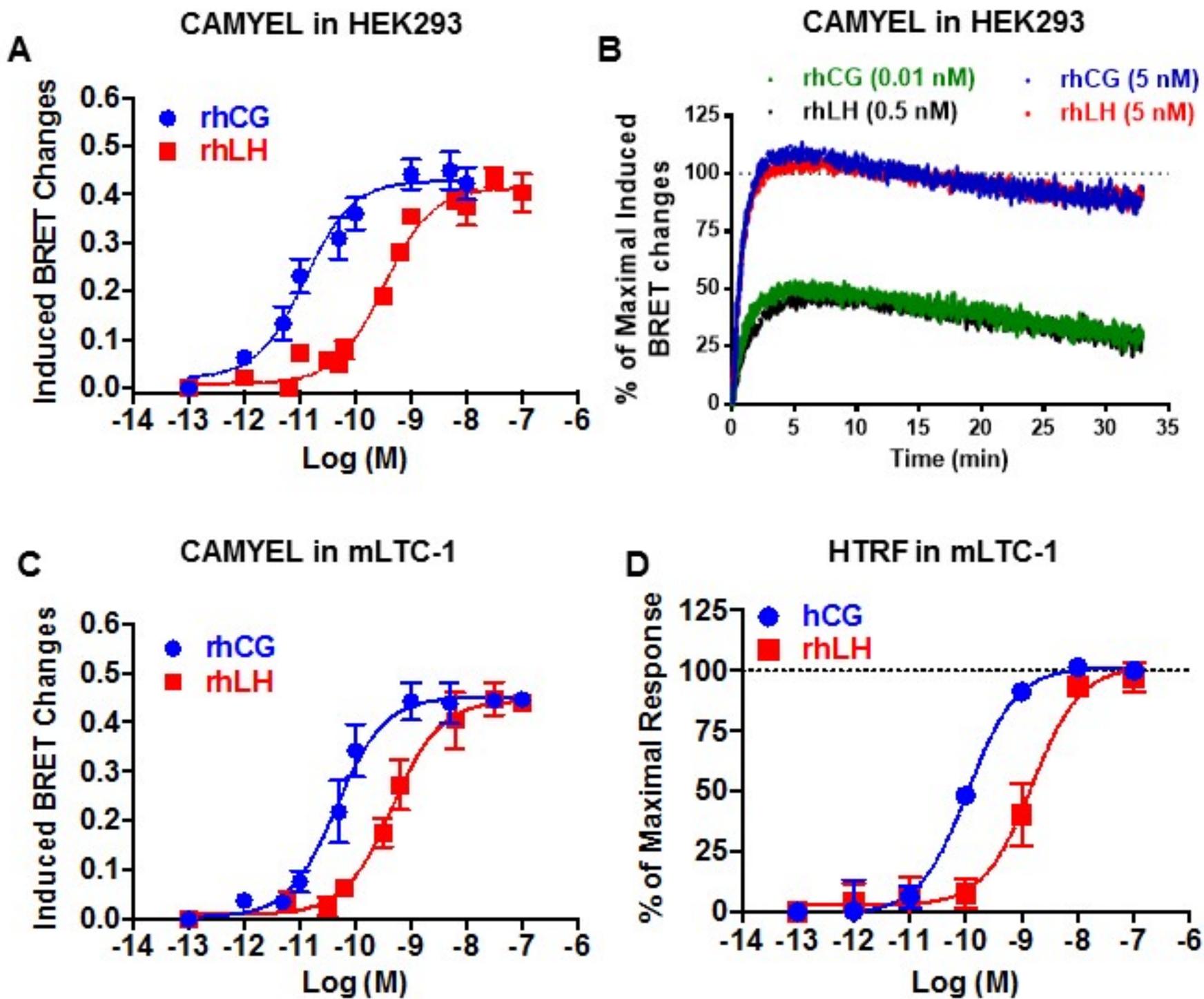

# Figure 2

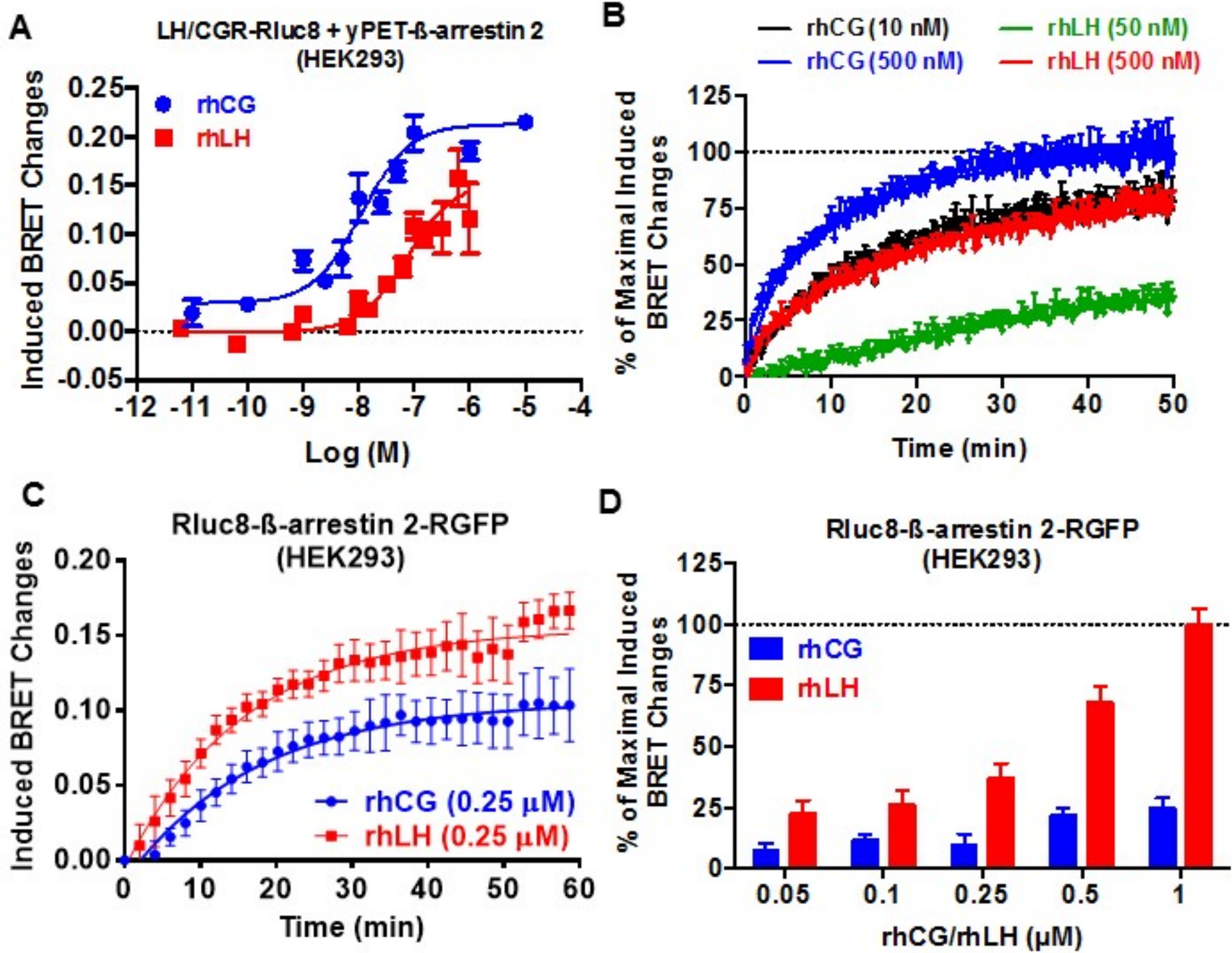

# Figure 3

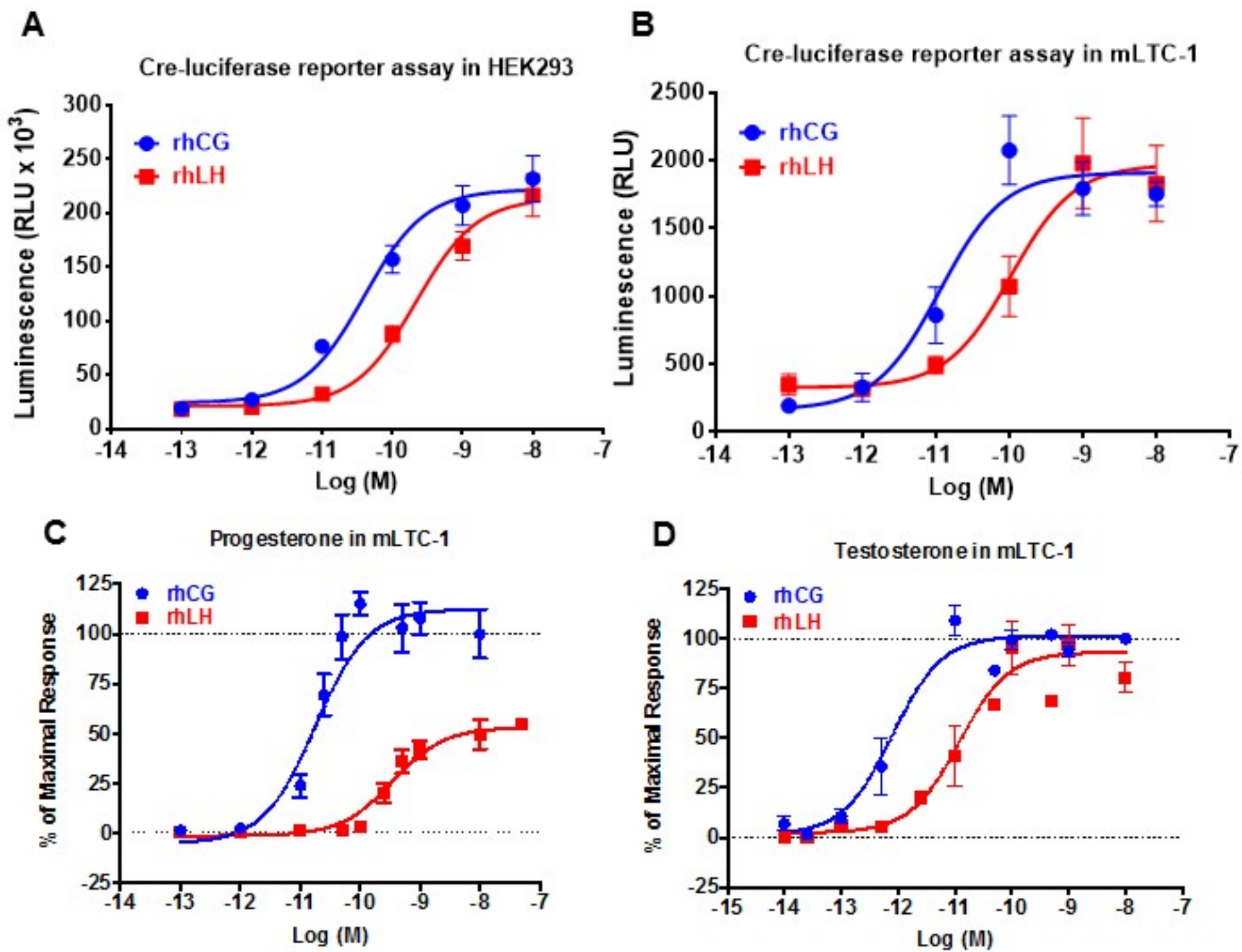

# Figure 4

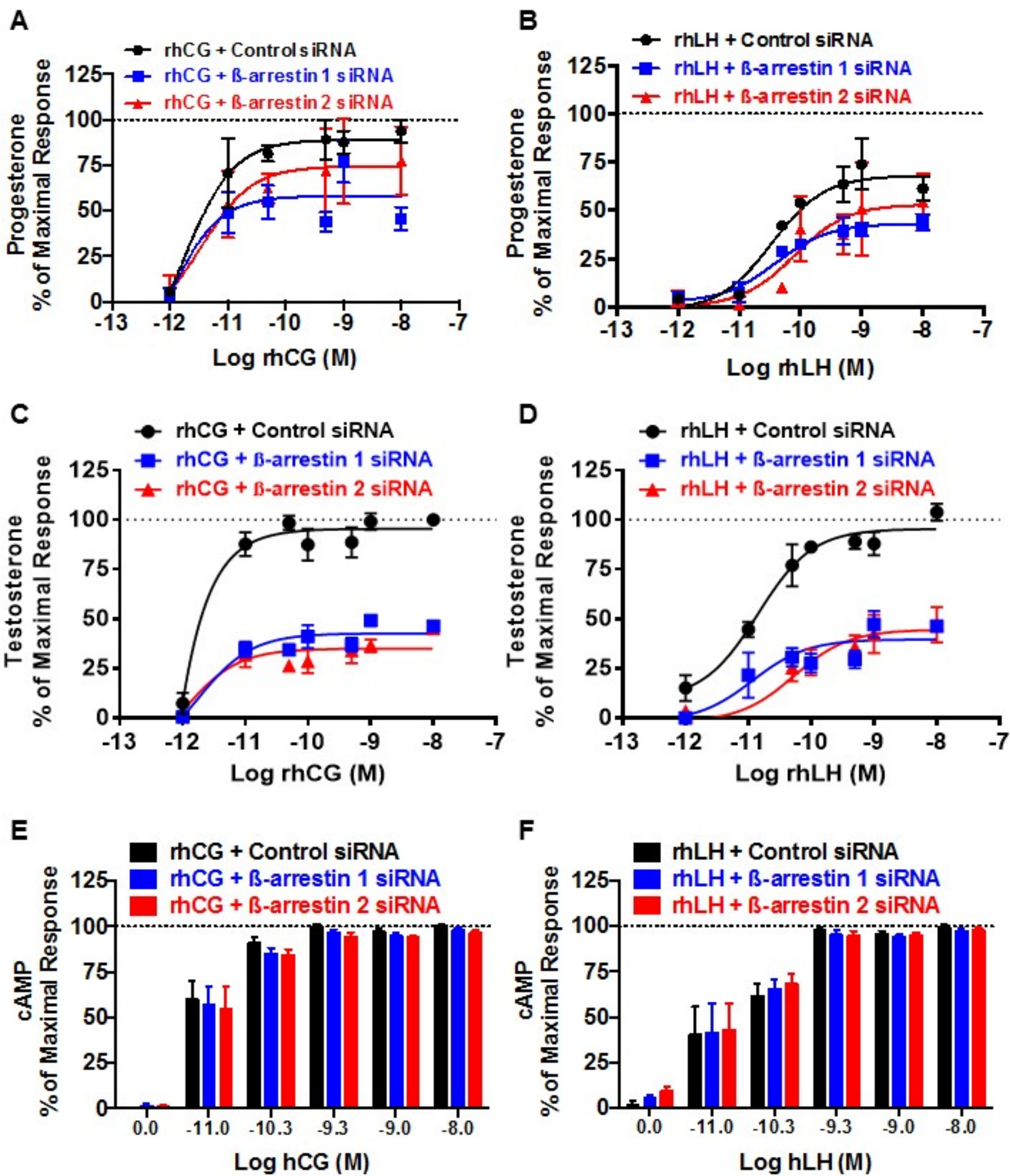